\documentclass[%
reprint,
 amsmath,
 amssymb,
 noeprint,
 aps,
 prl,
floatfix,
]{revtex4-1}

\usepackage{graphicx}% Include figure files
\usepackage{dcolumn}% Align table columns on decimal point
\usepackage{bm}% bold math
\newlength{\onecolfig}
\newlength{\twocolfig}
\usepackage{physics}
\usepackage{siunitx}

\makeatletter
\renewcommand*{\@fnsymbol}[1]{\ensuremath{\ifcase#1\or \dagger\or *\or \ddagger\or
    \mathsection\or \mathparagraph\or \|\or **\or \dagger\dagger
    \or \ddagger\ddagger \else\@ctrerr\fi}}
\makeatother

\newcommand{\dg}{\ensuremath{^\circ}}

\newcommand{\G}{\,\mathrm{G}}

\newcommand{\kHz}{\,\mathrm{kHz}}
\newcommand{\MHz}{\,\mathrm{MHz}}
\newcommand{\GHz}{\,\mathrm{GHz}}
\newcommand{\THz}{\,\mathrm{THz}}
\newcommand{\um}{\,\mathrm{\mu m}}

\newcommand{\nm}{\,\mathrm{nm}}

\newcommand{\us}{\,\mathrm{\mu s}}
\newcommand{\ms}{\,\mathrm{ms}}

\newcommand{\mW}{\,\mathrm{mW}}

\newcommand{\ct}{\ensuremath{^{43}\mathrm{Ca}^+}}
\newcommand{\sr}{\ensuremath{^{88}\mathrm{Sr}^+}}
\newcommand{\ddKet}{\ensuremath{\ket{\Downarrow \downarrow}}\,}
\newcommand{\uuKet}{\ensuremath{\ket{\Uparrow\uparrow}}\,}
\newcommand{\duKet}{\ensuremath{\ket{\Downarrow\uparrow}}\,}
\newcommand{\udKet}{\ensuremath{\ket{\Uparrow \downarrow}}\,}

\newcommand{\up}{\ensuremath{\ket{\uparrow}}\,}
\newcommand{\dn}{\ensuremath{\ket{\downarrow}}\,}
\newcommand{\Up}{\ensuremath{\ket{\Uparrow}}\,}
\newcommand{\Dn}{\ensuremath{\ket{\Downarrow}}\,}

\newcommand{\psz}{\ensuremath{\sigma_z}}

\newcommand{\OUp}{\ensuremath{\Omega_\Uparrow}}
\newcommand{\ODn}{\ensuremath{\Omega_\Downarrow}}
\begin{document}

\preprint{APS/123-QED}

\title{Benchmarking a high-fidelity mixed-species entangling gate}

\author{A.\,C.\,Hughes}
\thanks{These authors contributed equally.}
\author{V.\,M.\,Sch\"afer$^\dagger$}
\email{vera.schafer@physics.ox.ac.uk}
\author{K.\,Thirumalai}
\thanks{These authors contributed equally.}
\author{D.\,P.\,Nadlinger}
\author{S.\,R.\,Woodrow}
\author{D.\,M.\,Lucas}
\author{C.\,J.\,Ballance}
\affiliation{Department of Physics, University of Oxford, Clarendon Laboratory, Parks Road, Oxford OX1 3PU, U.K.}

\date{31 July 2020}

\begin{abstract}
We implement a two-qubit logic gate between a \ct\,hyperfine qubit and a \sr\,Zeeman qubit. 
For this pair of ion species, the S--P optical transitions are close enough that a single laser of wavelength $402\nm$ can be used to drive the gate, but sufficiently well separated to give good spectral isolation and low photon scattering errors.
We characterize the gate by full randomized benchmarking, gate set tomography and Bell state analysis. 
The latter method gives a fidelity of $99.8(1)\%$, comparable to that of the best same-species gates and consistent with known sources of error.
\end{abstract}

\maketitle

The exchange of quantum information between different types of qubit is a powerful tool for a wide variety of applications: these range from quantum computing and networking, to optical clocks, to spectroscopy of molecules or exotic species for testing fundamental physics.
In quantum information processing (QIP), interfacing different systems allows the use of specialized qubits for different operations (for example, memory, logic and readout) \cite{Wineland1998,Pfaff2013}, coupling of matter and photonic qubits for communication or distributed computing \cite{Wallraff2004,Ritter2012,Hucul2015,Nigmatullin2016a,Mi2017,Humphreys2018,Stephenson2020}, and even the connection of distinct qubit platforms (such as solid state and atomic systems \cite{Meyer2015,Petersson2012,Marcos2010}). 
While incoherent quantum state transfer often suffices for spectroscopy or clocks \cite{Schmidt2005,Chou2017,Brewer2019,Schmoeger2015,Smorra2015}, general QIP requires two-qubit interactions that preserve phase and amplitude of superposition states, i.e.\ entangling gate operations. 

Trapped-ion systems are an extremely promising technology for QIP, further enriched by the diversity of atomic properties of different species.
The use of two spectrally-resolved species allows laser cooling, state preparation and readout via one species without corrupting logic qubits held in the second species \cite{Barrett2003,Home2009}.
Different species also facilitate networking, where ions whose level structures and transition wavelengths are well suited to photonic interfacing can be gated with ions possessing superior properties for logic or memory \cite{Nigmatullin2016a,Inlek2017}.
Both mixed-isotope and mixed-element two-qubit gates have previously been demonstrated \cite{Ballance2015,Tan2015,Inlek2017,Negnevitsky2018,Bruzewicz2019,Wan2019}. 
However, due to the extra technical complications and new sources of error to which multi-species gates are susceptible, the gate fidelities achieved have fallen short of the state-of-the-art for single-species gates \cite{Ballance2016,Gaebler2016,Schafer2018,Erhard2019,Baldwin2020}. 

In this work, we perform a mixed-element $\psz\otimes\psz$ geometric phase gate \cite{Leibfried2003} between \ct\ and \sr, where the ions are driven by a state-dependent force tuned close to resonance with a motional mode. 
We take advantage of the extremely low addressing errors ($<10^{-6}$) of qubits with different energy splittings to characterize the gate performance using randomized benchmarking (RBM) with full exploration of the two-qubit Hilbert space, by gate set tomography (GST), and by partial Bell state tomography (PST). 

\begin{figure}
\centering
\includegraphics[width=\onecolfig]{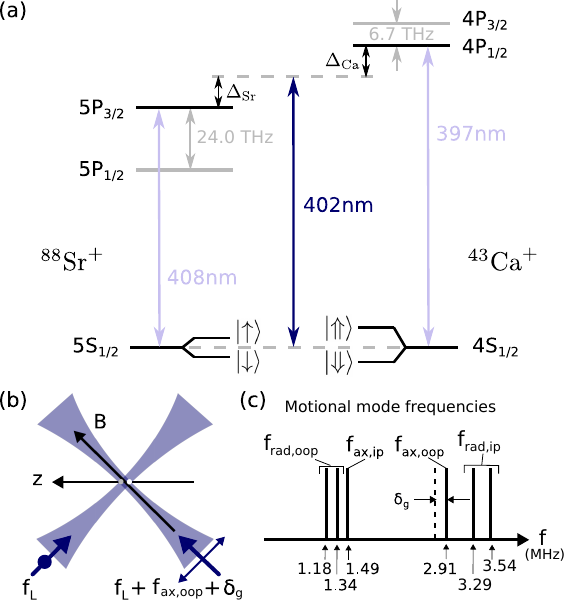}
\caption{ (a) A single laser with wavelength $\lambda=c/f_L=402\nm$, detuned by 
$\Delta_{\mathrm{Sr}} = +11.2\THz$, $\Delta_{\mathrm{Ca}}= -9.0\THz$ from S--P transitions in \sr\ and \ct, provides qubit-state-dependent forces on both ion species.
The qubit states are ($\dn\!,\up$)=($5\mathrm{S}_{1/2}^{-1/2}, 5\mathrm{S}_{1/2}^{+1/2}$) in \sr\ and ($\Dn\!,\Up$)=($4\mathrm{S}_{1/2}^{4,+4},4\mathrm{S}_{1/2}^{3+3}$) in \ct.
A static magnetic field $B=146\G$ gives qubit frequencies $f_\updownarrow=409\MHz$ and $f_\Updownarrow=2.874\GHz$. 
(b) Both Raman beams are derived from a single frequency-doubled Ti:S laser. 
The beatnote ($f_{\mathrm{ax,oop}}+\delta_g$) between the two beams is created by acousto-optic modulators. 
The differential wave vector ${\Delta\bm k}$ of the Raman beams is parallel to the trap axis $\bm z$, suppressing coupling to the radial modes of motion. 
(c) Values of (ax)ial and (rad)ial in-phase (ip) and out-of-phase (oop) normal mode frequencies.}
\label{fig:setup}
\end{figure}

Calcium and strontium are a particularly well-matched pair of elements: 
they have S--P transitions at $397\nm$ and $408\nm$, with linewidths $\Gamma\approx22\MHz$, which are only $\Delta_0=20.2\THz$ apart (Fig.\,\ref{fig:setup}a). 
This frequency separation is small enough that a single laser of modest intensity, at $\lambda\approx402\nm$, can provide a state-dependent force on both species simultaneously to perform a two-qubit gate. Nevertheless, $\Delta_0$ is large enough to give small ($\sim10^{-4}$) photon scattering errors during the gate, and to preserve excellent spectral isolation during cooling and readout, as ($\Gamma/\Delta_0)^2\sim10^{-12}$ \cite{Ballance2014,Schafer2018a}.
The Raman beatnote required to drive a $\psz\otimes\psz$ gate depends only on the motional mode frequency and is independent of the qubit frequency.
Therefore a single pair of Raman beams can perform a gate between strontium and calcium ions, in a  similar manner to the gate previously performed on two different isotopes of calcium \cite{Ballance2015}.
This simplifies the technical setup compared with previous demonstrations of mixed-element logic gates, which required separate laser systems for each species \cite{Tan2015,Inlek2017}. 
The masses of \ct\ and \sr\ differ only by a factor of two, yielding sufficient motional coupling for efficient sympathetic cooling \cite{Morigi2001,Home2013}. 
\ct\ has proved to be an excellent logic qubit, with stable hyperfine states allowing single-qubit gates and memory with errors $\sim10^{-6}$ \cite{Harty2014,Sepiol2019}, and state preparation, readout and two-qubit gates with errors $\lesssim10^{-3}$ \cite{Harty2014,Harty2016,Ballance2016}. 
\sr\ is superior for optical networking purposes due to its simple level structure and favourable wavelengths: 
high-rate, high-fidelity entanglement of ions in separate vacuum systems via a photonic link has recently been demonstrated with \sr\ ions \cite{Stephenson2020}. 

Photon scattering sets a quasi-fundamental limit to the error for two-qubit gates driven by stimulated Raman transitions, and was the largest error contribution in the highest-fidelity two-qubit gates \cite{Ballance2016,Gaebler2016}. 
To achieve a scattering error $\sim10^{-4}$ requires several-THz detuning $\Delta$ of the Raman beams from resonance \cite{Ozeri2007}.
The calculated scattering error for a laser tuned midway between the \ct\ and \sr\ S--P transitions is $2\times 10^{-4}$, about a factor two lower than that in \cite{Ballance2016,Gaebler2016}. 
The dominant technical errors are often due to motional and spin dephasing of the ions, which are smaller the faster the gate \cite{Ballance2016,Schafer2018}. 
To maintain a reasonable gate speed at THz detunings requires a beam power of a few tens of mW for $30\um$ spot sizes. 
We choose the Raman detuning to be roughly half-way between the nearest S--P transitions, at $\Delta_{\mathrm{Ca}}=-9.0\THz$ from the $397\nm$ transition in \ct, to couple similarly to both species. For our setup, this detuning also approximately minimizes the error due to motional heating during the gate \cite{Schafer2018a}.

\begin{figure}
\centering
\includegraphics[width=\onecolfig]{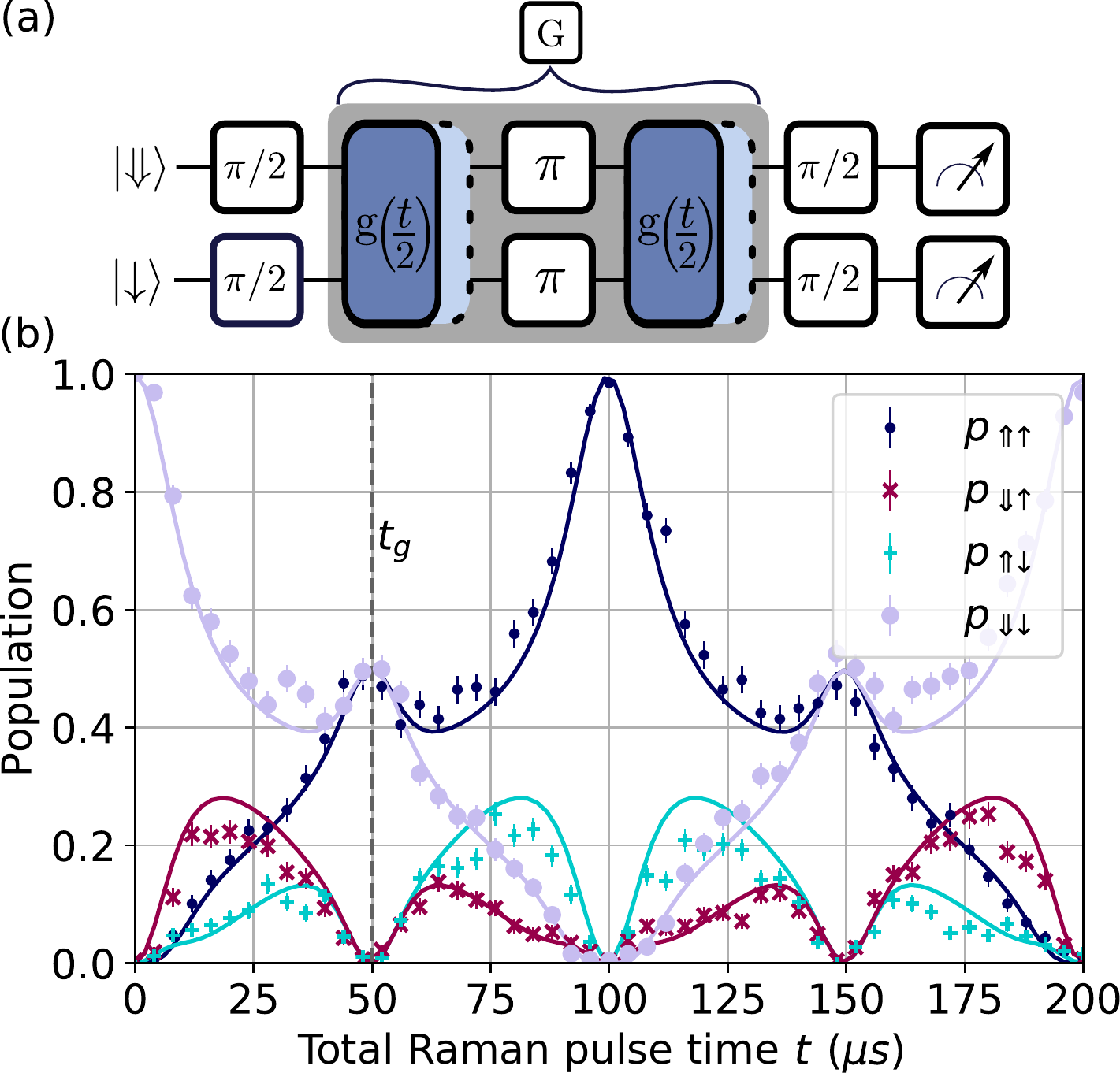}
\caption{(a) Schematic sequence of the $\psz\otimes\psz$ gate. 
The gate $G$ characterized in RBM and GST measurements consists of two Raman pulses $\mathrm{g}(t_{\mathrm{g}}/2)$, separated by a ``spin-echo'' $\pi$-pulse on each qubit.
For PST, $G$ is bracketed by two $\pi/2$-pulses, and followed by an additional $(\pi/2)_\phi$ analysis pulse (not shown), with phase $\phi=45\dg$ or $135\dg$, to determine the parity contrast. 
(b) Gate dynamics measured with the sequence in (a). 
To improve sensitivity during calibration, the initial phase of the second gate pulse matches that of the first pulse.
The asymmetry of the forces on \ct\ and \sr\ leads to asymmetric evolution of the populations $p_{\Downarrow \uparrow }$, $p_{\Uparrow \downarrow}$.}
\label{fig:gate_dyn}
\end{figure} 

\begin{figure*}
\includegraphics[width=59mm]{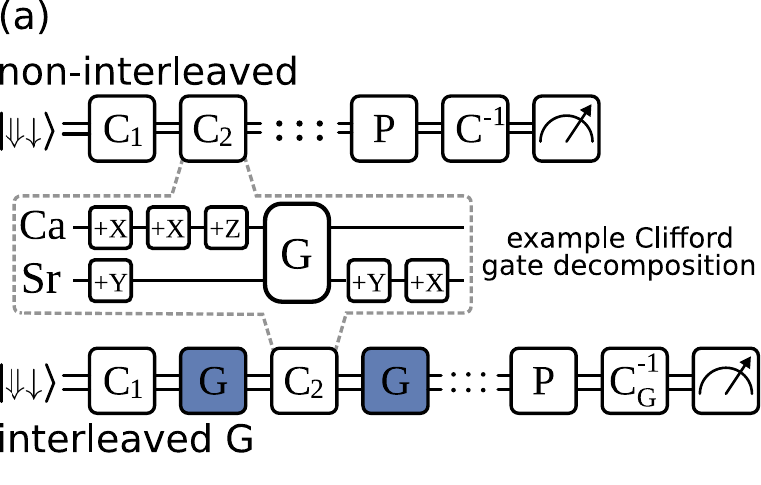} 
\includegraphics[width=119mm]{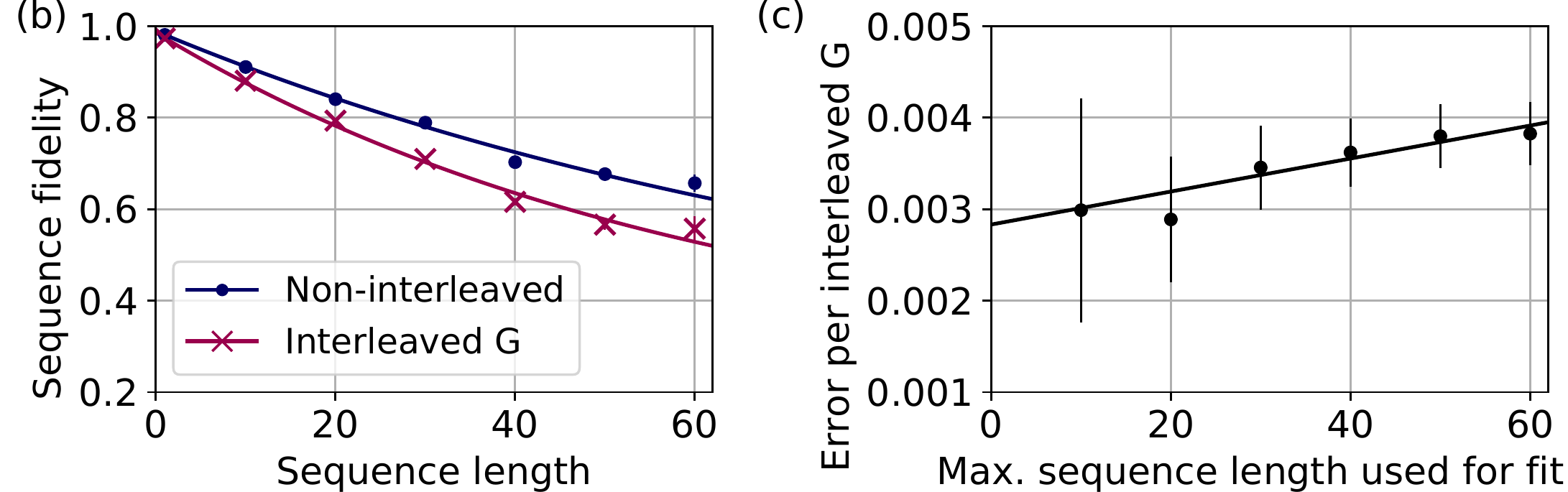}  
\caption{Interleaved RBM: 
(a) Gate sequences. Operations $C_1\ldots C_L$ are randomly chosen from the 11520 elements of the two-qubit Clifford group, generated by the single-qubit $\pi/2$-rotations $\{\pm{X},\pm{Y},\pm{Z}\}$, and $G$. In interleaved sequences, the entangling gate-under-test $G$ is applied after each $C_i$. To finish in an eigenstate of the measurement basis, each sequence terminates with the inverse operation $C^{-1}$ ($C^{-1}_G$) of all combined $C_i$ ($C_i$ and $G$) operations. The final eigenstate is randomized by single-qubit $\pi$-rotations $P$.
(b) Combined datasets of 660 randomized sequences with $L=1\ldots60$. Each randomization is repeated 100 times to determine the sequence fidelity. Each datapoint gives the mean fidelity for all sequences of length $L$.
(c) Error per interleaved $G$ gate, inferred from fitting the fidelity decays in (b), as a function of the maximum sequence length used in the fit. The error-per-gate increases with $L$ as certain error sources become more pronounced after longer durations. Error bars are from parametric bootstrapping.
}
\label{fig:rbm}
\end{figure*} 

The gate is implemented as two separate pulses embedded in a spin-echo sequence (Fig.\,\ref{fig:gate_dyn}a), each pulse driving a closed-loop trajectory in motional phase space.
This two-loop scheme reduces spin decoherence and cancels effects of the asymmetry of \OUp{} and \ODn{} in \ct\ \cite{Ballance2014}, where the Rabi frequencies \OUp{} and \ODn{} determine the strength of the light-shift force on the two different qubit states.
The phase of the second gate pulse is flipped by $\pi$ compared to the gate dynamics in Fig.\,\ref{fig:gate_dyn}b, to perform a first-order Walsh modulation which reduces sensitivity to mis-set parameters \cite{Hayes2012}.
In the $\psz\otimes\psz$ gate, the phase of the single-qubit operations needs no fixed relationship to the Raman beatnote phase \cite{Lee2005}.
Single-qubit operations may therefore be driven independently using microwaves, with no phase coherence to the gate beams.
The gate is performed on the axial out-of-phase (oop) motional mode, with a gate detuning $\delta_g=-40\kHz$ giving a gate time $t_g=2/\delta_g=49.2\us$.
The edges of the pulse are shaped by a Hann window, $\mathrm{sin}^2(\pi t/2t_\mathrm{s})$, with $t_\mathrm{s}=2\us$. 
The power in each Raman beam is $60\mW$, in a spot size $\approx 30\um$, giving $\ODn/2\pi=180\kHz$.
For maximum gate efficiency, i.e.\ maximum two-qubit geometric phase acquired for a given carrier Rabi frequency, the $\psz\otimes\psz$ gate requires the ion spacing to be an integer or half-integer multiple of the Raman beam standing wave period $\lambda_z=402\nm/\sqrt{2}$. 
The ion spacing is $12.5\lambda_z=3.57\um$, corresponding to axial mode frequencies shown in Fig.\,\ref{fig:setup}c. 
The axial Lamb-Dicke parameters $(\eta_\mathrm{ip}, \eta_\mathrm{oop})$ are $(0.090, 0.127)$ for \ct\ and $(0.124,0.045)$ for \sr. 
The trap frequencies and the sign of $\delta_g$ need to be chosen carefully to avoid several resonances: 
(i) $f_{\mathrm{ax,oop}}\simeq 2f_{\mathrm{rad,oop}}$; 
(ii) $2f_\mathrm{ax,ip}\simeq f_\mathrm{ax,oop}+\delta_g$; and, for driving an in-phase (ip) mode gate,
(iii) $f_{\mathrm{ax,ip}}\simeq f_{\mathrm{rad,oop}}$. 
The axial modes were sub-Doppler cooled but the radial modes were not. 
For case (i), hot radial modes lead to large errors due to Kerr cross-coupling.
For (ii), certain $\delta_g$ can lead to errors from higher harmonic excitation \cite{Bruzewicz2019}.
For (iii), and if the Raman differential wave vector $\Delta\bm k$ is not orthogonal to the radial modes, additional errors can arise from radial mode excitation \cite{Schafer2018}. 

We characterize the error $\epsilon_G$ of the gate operation using three different methods:
by measurement of the Bell state fidelity after a single gate operation using PST, and in longer sequences of gates using interleaved RBM and GST. 
For PST with a single gate operation we obtain $\epsilon_G=2.0(1.0)\times10^{-3}$ after correction for state preparation and measurement (SPAM) errors \cite{Ballance2015}, averaged over 50~000 gate measurement shots taken over two separate days~\footnote{We also performed the gate on the axial in-phase mode, obtaining $\epsilon_G=2.2(9)\times10^{-3}$ by PST~\cite{Thirumalai2019}.}.
The raw gate error before SPAM correction is $10.2(3)\times10^{-3}$.
Errors due to other single-qubit operations were not corrected for.
For readout, the \sr\ ion is shelved with a $674\nm$ pulse, which is sensitive to the ion temperature; for \ct\ a $393\nm$ optical pumping process is used, which is far less temperature dependent \cite{Myerson2008}. 
As the ions are heated to Doppler temperature during fluorescence detection, we read out \sr\ first.
The average SPAM errors are $\bar{\epsilon}_{\mathrm{Sr}}=4.0(3)\times10^{-3}$ for \sr\ and $\bar\epsilon_{\mathrm{Ca}}=1.4(3)\times10^{-3}$ for \ct, and the dominant uncertainty in $\epsilon_G$ arises from statistical uncertainty in $\bar{\epsilon}$.

RBM allows measurement of the gate error independent of SPAM errors, and also provides a measure of the gate error in the more computationally relevant context of long sequences of operations \cite{Knill2008,Gaebler2012,Baldwin2020}. 
We implement an interleaved RBM protocol equivalent to \cite{Gaebler2012}, with the $\psz\otimes\psz$ operation as the ``gate-under-test'' $G$. 
The error per $\psz\otimes\psz$ gate $\epsilon_G$ is extracted by comparing the error rate $\epsilon_g$ of sequences of random Clifford operations to the error rate $\epsilon_g'$ of sequences which also include extra interleaved $\psz\otimes\psz$ gates, via \cite{Gaebler2012}:
\begin{align}
\epsilon_G=\frac{1}{\alpha_n}\left[ 1-\frac{1-\alpha_n\epsilon_g'}{1-\alpha_n\epsilon_g} \right]
\end{align}
with depolarization probability $\alpha_n\epsilon$, where $\alpha_n=2^n/(2^n-1)$ and $n$ is the number of qubits. 
Individually-addressed single-qubit $X$ and $Y$ gates are achieved trivially through the use of different ion species, and single-qubit $Z$-rotations are implemented as software phase shifts. 
As both species participate in the gate operation, no sympathetic cooling is performed during the sequences. 
Sequences and results are shown in Fig.\,\ref{fig:rbm}. 
For each sequence length (up to 60 interleaved gates) we perform $N = 100$ shots of $k \simeq 100$ randomly-generated pairs of sequences.
Each two-qubit Clifford operation contains on average 1.5 entangling operations as well as 7.7 single-qubit rotations, meaning that the longest sequences contain on average 151.5 entangling operations in total, and take $16\ms$. 
Over these durations the increase in ion temperature and duty-cycle effects in amplifiers become non-negligible, leading to a slight increase in the error-per-gate for longer sequences (Fig.\,\ref{fig:rbm}c). 
The error due to ion temperature is dominated by the ip mode heating ($\approx110\,\mathrm{quanta/s}$): after $16\ms$, $\bar{n}_{\mathrm{ax,ip}}\approx1.8$, contributing $3\times 10^{-3}$ error for the last oop-mode gate in the sequence. 
We measure the average error-per-gate $\epsilon_G=2.9(7)\times10^{-3}$ ($3.8(3)\times10^{-3}$) from sequences with 20 (60) interleaved entangling gates-under-test, corresponding to sequences involving up to 51.5 (151.5) entangling gates.
From the non-interleaved sequences alone, we can also extract the average error of an arbitrary two-qubit Clifford operation $\epsilon_g = 8.3(2)\times10^{-3}$, consistent with the errors of its constituent operations.

\begin{figure*}
\begin{minipage}[b]{56mm}
\includegraphics[width=\textwidth]{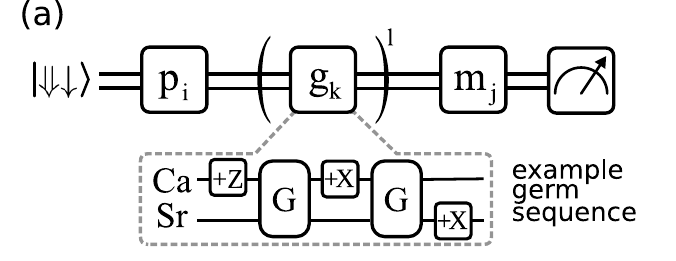}
\includegraphics[width=\textwidth]{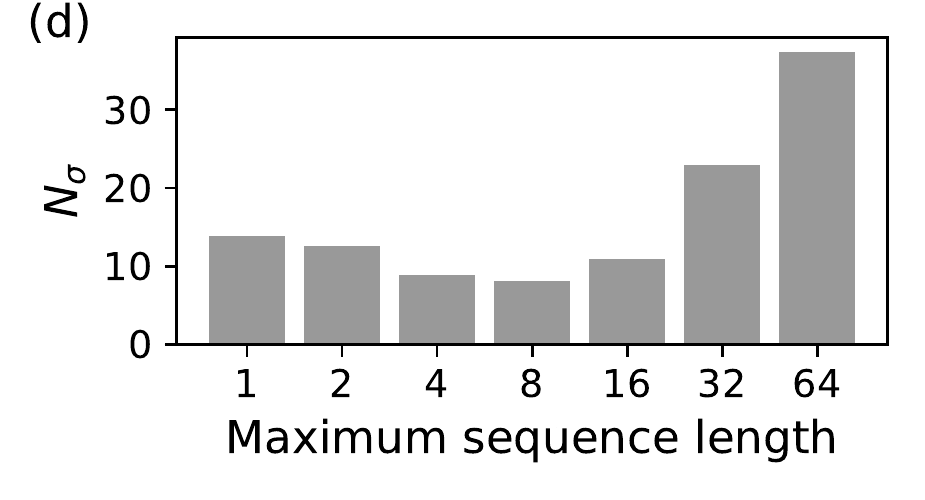}
\end{minipage}
\hspace{10mm}
\includegraphics[width=50mm]{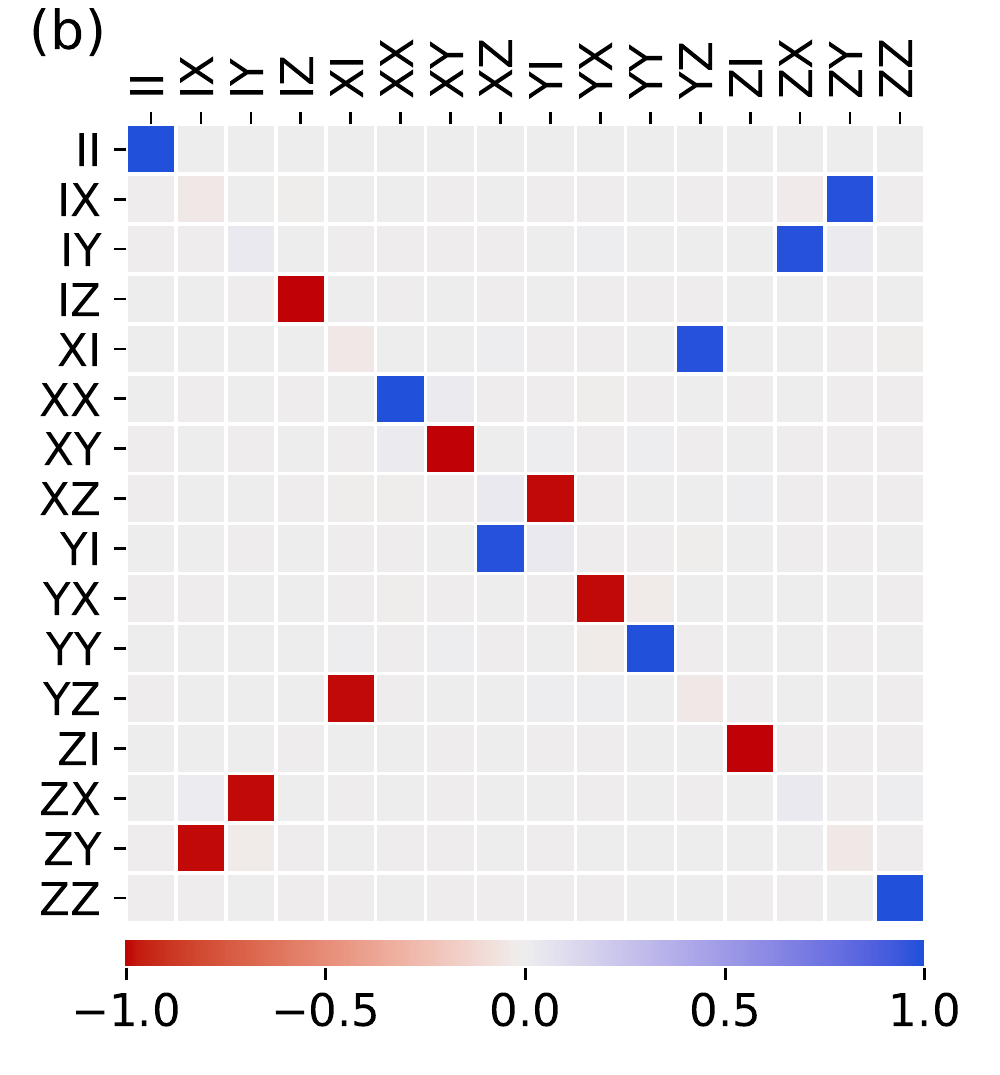}
\hspace{5mm}
\includegraphics[width=50mm]{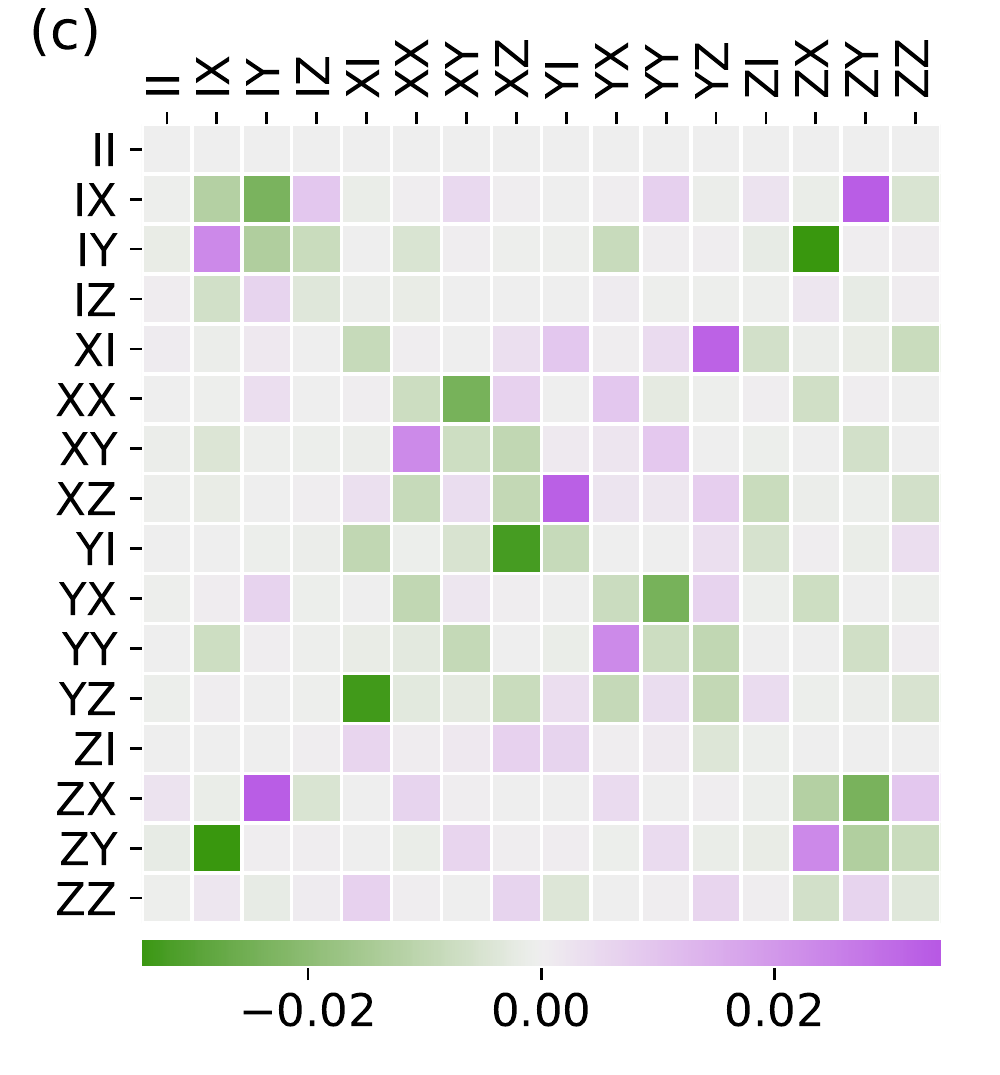}
\caption{Gate set tomography: (a) Structure of GST sequences. 
For each target sequence length $L = 1, 2, 4, \ldots, 64$, each germ sequence $g_k$ is repeated $l$ times, such that the maximum number of operations is $L$, and is surrounded by combinations of preparation and measurement fiducials $p_i$ and $m_j$. 
(b) Final maximum-likelihood estimate for the entangling gate under CPTP constraints, given as a superoperator $\mathcal{G}$ in the Pauli-product basis. 
(c) Difference of the fit of the measured gate $\mathcal{G}$ to the ideal gate $\mathcal{G}_0$.
Displayed is the error generator $\mathbb{L}$ such that $\mathcal{G} = \mathrm{e}^\mathbb{L}\, \mathcal{G}_0$. 
(d) Goodness-of-fit of the final estimate for data up to circuit length $L$, in units of standard deviations of the expected $\chi^2$ distribution. 
A significant amount of non-Markovianity is present, presumably due to duty-cycle effects and motional heating.
}
\label{fig:gst}
\end{figure*}

To obtain a measure of the nature of the errors of different operations, in addition to their magnitude, GST can be employed \cite{Blume-Kohout2013}.
We characterize the gate set $\left\{G, +X, +Z\right\}$, which contains 1026 gauge-independent parameters.
To reduce the time necessary to estimate individual errors, we implement repetitions of shorter gate sequences, called germs, which are chosen to amplify specific errors \cite{Blume2017}.
A total of 62 germ sequences are used, each consisting of 1 to 5 operations. 
The germ sequences are optimized numerically, in turn amplifying sensitivity to a specific noise parameter and minimizing the number of required operations.
The germ sequences are surrounded by a preparation and a measurement fiducial, Fig.\,\ref{fig:gst}a.
The fiducials serve to form complete sets of prepared states and measurement effects akin to process tomography.
They are chosen from a set of 15 preparation and 10 measurement fiducials, each consisting of 1 to 4 operations per qubit.
Operation sequences are constructed and analyzed from these sets using the pyGSTi package \cite{pyGSTI}.
The error of the entangling gate $G$ extracted from the results (Fig.\,\ref{fig:gst}) is $\epsilon_G=6(3)\times10^{-3}$. 
The diamond norm distance $||.||_{\diamond}=0.03(1)$, of which half is due to coherent error sources.

The three gate characterization methods have different merits. 
For the above results, the total number of qubit population measurements and the time for a single measurement, including cooling, are: ($250\,000$, $16\ms$) for PST including SPAM measurements; ($132\,000$, $16\ms$--$32\ms$) for interleaved RBM; ($386\,000$, $16\ms$--$23\ms$) for GST. 
PST is the simplest to implement, as it does not require individual qubit addressing and long sequences.
The accuracy of RBM and GST is limited by the increase of gate error with sequence length, which is not included as a fit parameter, and therefore reduces the quality of the fit.
While PST is not affected by this systematic effect, it requires more data than RBM to achieve equal statistical uncertainties, and can be subject to systematic error due to drifts in SPAM errors.
GST provides information about the nature of gate errors that is essential for choosing and designing appropriate error correction algorithms \cite{Gottesman2010}.
However the reduction of the accuracy versus data set size, due to the observed non-Markovianity (Fig.\,\ref{fig:gst}d), as well as the computational complexity of the analysis, make GST less convenient.
In our regime of comparable gate and SPAM errors, RBM provides the best accuracy for a given data acquisition time, and also reveals effects occurring in longer sequences.

The operation of mixed-species gates is qualitatively different, and has increased sensitivity to certain experimental imperfections, compared with same-species gates.
Owing to the different masses and atomic structure, the sideband Rabi frequencies for the two ions are different; thus the light-shift force $F_\uparrow\neq F_\Uparrow$ (and $F_\downarrow\neq F_\Downarrow$) and, in general, all four states (\uuKet\!,\udKet\!,\duKet\!,\ddKet\!) are motionally excited. 
% -1 from oop mode, -1 from half-integer ion spacing -> even parity excited
Hence some (here $\approx$~20\%) of the acquired geometric phase gives rise to a global phase  instead of to the desired two-qubit phase generated by $\psz\otimes\psz$. 
Therefore the total Rabi frequency has to be increased (by $\approx$~3\%) and the maximum excursion in phase space becomes larger. 
This increases errors due to heating, photon scattering, and imperfect closure of loops in phase space, compared with more efficient single-species gates.
Owing to the mass-dependence of the rf pseudo-potential, mixed-species crystals tilt off the trap $z$-axis if there are stray electric fields that displace the ions radially \cite{Home2013}.
This leads to errors, for example by increasing the coupling of the gate beams to radial modes, or by increasing sensitivity to noise in the rf drive.
Empirically, in the presence of a stray radial field, we measure gate errors much greater than for Ca--Ca crystals, and larger than can be accounted for by a simple model of the effect of crystal tilt. 
We therefore adjust the field compensation more precisely than for a single-species crystal and, to mitigate the effects of trap field inhomogeneities \cite{Home2013}, we keep the ion order constant \cite{Home2011,Schafer2018a}. 
By measuring the quadratic dependence of the gate error on the stray field, we compensate the field to $\approx 0.3\,\mathrm{V}/\mathrm{m}$ precision; the effect of this change in field on the gate error was measured by RBM to be $2(5)\times 10^{-4}$.

\begin{table}
\begin{tabular}{lr} \hline \hline
\multicolumn{2}{c}{{Error source}\hfill{Error $(\times10^{-4})$}} \\ \hline\hline
Single-qubit $\pi/2$ rotations (measured by RBM)											& $4.3(2)$ \\ \hline
Imperfect stray field compensation (measured)												& $<7$ \\
Ion heating during gate ($\dot{\bar{n}}_\mathrm{ax,oop}\approx30\,\mathrm{quanta/s}$) 	& $4$ \\
Raman + Rayleigh photon scattering 															& $2$ \\
Kerr cross-coupling (assume similar to same-species) 									& $\lesssim 2$ \\
Coupling to spectator modes 																		& $1$ \\
Spin dephasing 																						& $<1$ \\ \hline\hline
\end{tabular}
\caption{Error contributions for the Ca--Sr gate. Single-qubit rotations contribute differently to the three methods due to the varied number of $\pi$ and $\pi/2$ pulses. Unless otherwise indicated, these values are calculated (see \cite{Ballance2014,Schafer2018a}).}
\label{T:errorbudget}
\end{table}

In summary, we have performed a mixed-element entangling gate between \ct\ and \sr\ ground-level qubits using a light-shift gate previously used for same-element quantum logic. 
We have demonstrated comparative benchmarking of the gate using three independent methods (PST, RBM and GST), which yield respective fidelities 99.8(1)\%, 99.7(1)\% and 99.4(3)\% (consistent at the $\approx\!\!1\sigma$ level). 
The PST fidelity is consistent with that of the best same-species gates, as measured by the same method \cite{Ballance2016,Gaebler2016,Harty2016,Schafer2018,Zarantonello2019,Srinivas2020}, and with the total error from known sources (Table \ref{T:errorbudget}). 
The gate mechanism requires one pair of beams from a single, visible-wavelength c.w.\ laser, near the technically-convenient $405\nm$ band, making it promising for use in scalable trapped-ion quantum computing architectures.

\vspace{1ex}

% Acknowledgements

We thank R.~Srinivas for comments on the manuscript. VMS is a Junior Research Fellow at Christ Church, Oxford. SRW was funded by the U.K.\ National Physical Laboratory, KT by the U.K.\ Defence Science and Technology Laboratory. CJB is supported by a UKRI FL Fellowship, and is a Director of Oxford Ionics Ltd. This work was funded by the U.K.\ EPSRC ``Networked Quantum Information Technology'' and ``Quantum Computing and Simulation'' Hubs, and by the E.U.\ Quantum Technology Flagship project AQTION (820495).

%\bibliography{bibliography}% Produces the bibliography via BibTeX.

%Control: production of eprint (1) enabled
%

\end{document}